\documentclass[pre,amsmath,amssymb,11pt,nofootinbib,showpacs]{revtex4}
\def\pfoots{}\normalsize
\def\pnlabel#1{\hbox{\tiny[#1]}\label{#1}}
\usepackage{pnrevtex} 

\def\pnlabel#1{\label{#1}}

\usepackage{graphicx}
\usepackage{bm}
\def\pnmemo#1{}

\def\xibval{50}\def\xitval{18}\def\xitbulkval{70}\def\pthetaaval{71}
 \def\pavalue{4.0}
\newcommand{\dn}{\mathrm{dn}}
\newcommand{\sn}{\mathrm{sn}}
\newcommand{\cn}{\mathrm{cn}}
\newcommand{\am}{\mathrm{am}}
\newcommand{\Lk}{\mathrm{Lk}}
\newcommand{\Tw}{\mathrm{Tw}}
\newcommand{\Wr}{\mathrm{Wr}}
\def\pz{\zeta}
\def\st{_{\star}}
\def\pspacing{L}
\def\pLtot{L\tot}
\def\ptorq{M_{\rm ext}}
\def\pavec{\pnvec a}
\def\pasca{a}
\def\pthetaa{\theta_a}
\def\uvec#1{\hat{\pnvec #1}}
\def\pthat{\hat{\pnvec t}}
\def\dhat{\hat{\pnvec d}}
\def\ehat{\hat{\pnvec e}}
\def\pini{_{\rm i}}
\def\pfin{_{\rm f}}
\def\plac{\textit{lac}}
\def\puns{_{\rm un}}

\def\starc{$\star$-coupler}
\def\st{_{\star}}
\def\pono{\omega_0}
\def\phelix{\pspacing_{\rm helix}}
\def\pktw{K_{\rm tw}}\def\pxitw{\xi_{\rm tw}}
\def\pkb{K_{\rm bend}}\def\pxib{\xi_{\rm b}}
\def\plst{\ploopn}\def\pulst{{\cal U}}
\def\bkap{\mbox{\boldmath$\kappa$}}
\def\pE{{\cal E}}
\def\ploopn{n}
\def\ppsi{\psi}

\newcommand{\lfr}[2]{{\textstyle\frac{#1}{#2}}}

\def\eref#1{Eq.~\ref{#1}}
\def\erefs#1{Eqs.~\ref{#1}} 
\def\sref#1{Sect.~\ref{#1}}\def\srefs#1{Sects.~\ref{#1}}
\def\fref#1{Fig.~\ref{#1}}
\def\frefn#1#2{Fig.~\ref{#1}{\it #2}}
\def\rref#1{Ref.~\cite{#1}}
\def\rrefs#1{Refs.~\cite{#1}}
\newcommand{\dd}{{\mathrm d}}
\newcommand{\pnvec}{\mathbf}         
\newcommand{\inv}{^{\raise.15ex\hbox{${\scriptscriptstyle -}$}\kern-.05em 1}}

\newcommand{\eff}{_{\rm eff}}

\def\kbt{{k_{\rm B}T}}
\def\tot{_{\mathrm{tot}}}

\def\nmunit{\ensuremath{\mathrm{nm}}}

\newcommand{\pitemize}{\begin{itemize}\setlength{\itemsep}{0pt}\setlength{\parsep}{0pt}}
\newcommand{\xpitemize}{\end{itemize}}
\newcommand{\penumerate}{\begin{enumerate}\setlength{\itemsep}{0pt}\setlength{\parsep}{0pt}}
\newcommand{\xpenumerate}{\end{enumerate}}

\def\ifigure#1#2#3#4{\begin{figure}
\begin{center}\includegraphics[height=#4truein]{#3}
\end{center}\smallskip 
\caption{\footnotesize #2 \pnlabel{#1}}
\end{figure}}
\def\ifigureab#1#2#3#4#5{\begin{figure}[tb!]
\hfil
\includegraphics[height=#5truein]{#3}\includegraphics[height=#5truein]{#4}
\hfil\par\smallskip \caption{\footnotesize #2 \pnlabel{#1}}
\end{figure}%
}

\begin{document}

\title{Effect of supercoiling on formation of protein mediated DNA loops}

\author{Prashant K. Purohit}
\email{purohit@seas.upenn.edu} \affiliation{Department of
Mechanical Engineering and Applied Mechanics, University of Pennsylvania,
Philadelphia, PA 19104, USA}

\author{Philip C. Nelson}
\email{nelson@physics.upenn.edu} \affiliation{Department of
Physics and Astronomy, University of Pennsylvania, Philadelphia PA
19104, USA}

\date{Resubmitted 13 Oct 06}
\begin{abstract}
DNA loop formation is one of several mechanisms used by organisms to
regulate genes. The free energy of forming a loop is an important
factor in determining whether the associated gene is switched on or
off. In this paper we use an elastic rod model of DNA to determine the
free energy of forming short (50--100 basepair), protein mediated DNA
loops. Superhelical stress in the DNA of living cells is a critical
factor determining the energetics of loop formation, and we explicitly
account for it in our calculations. The repressor protein itself is
regarded as a rigid coupler; its geometry enters the problem through
the boundary conditions it applies on the DNA. We show that a theory
with these ingredients is sufficient to explain certain features
observed in modulation of {\it in vivo} gene activity as a function of
the distance between operator sites for the {\it lac} repressor. We
also use our theory to make quantitative predictions for the dependence
of looping on superhelical stress, which may be testable both {\it in
vivo} and in single-molecule experiments such as the tethered particle
assay and the magnetic bead assay.
\end{abstract}

                                             
\maketitle

\section{Introduction and summary}
\subsection{Introduction}
Many genetic processes in bacteria are controlled by proteins that bind
at separate, often widely spaced, sites on DNA and hold the intervening
double helix in a loop \cite{dunn84a,adhy89a,schl92a,half04a}. For example, the
lactose metabolism system in \textit{E. coli} is controlled by a
repressor protein, LacI, binding to a set of binding sites. Early evidence for
the existence of a looping mechanism came from the
observation that the ability of a cell to control a
 particular gene
depended in an approximately periodic way upon the number of basepairs of DNA
intervening between two particular protein-binding sequences (called ``operators'')
(see for example
\cite{dunn84a,hoch86a,kram87a,bell88a}). Some recent data appear in 
\fref{ftwo}. The periodic modulation was found to be
roughly independent
of the details of what basepair sequence was inserted or deleted
between the operators; insertions and deletions elsewhere did not
affect the 
gene regulation in this way. 

\ifigure{ftwo}{\textit{a.} Repression of a gene product controlled by
 the \plac{} repressor in {\it E. coli} cells. 
The data are
 from \protect\cite{mull96a}, Figure 3a; see that paper for
an explanation of the units on the vertical axis. The horizontal axis 
gives the distance along the 
DNA between the centers of the two 
operators, each 21 basepairs long. In this paper we will express 
operator spacing by a number $\pspacing$ that equals this number minus 21bp 
(\fref{fone}).
Other experiments have obtained similar 
curves using operators located on a plasmid \protect\cite{beck05a}.
\textit{b.}~Looping free energy 
inferred from the data in (a), showing a roughly periodic modulation
with operator spacing  (from \protect\cite{saiz05a}, Figure 3). The
maxima of this function correspond to poor looping efficiency, that is,
to the minima in panel (a). There is a slight minimum in the lower
envelope of this function around 70--80bp, corresponding to our 
$\pspacing\approx50$--60bp. A similar function 
emerges from the more detailed analysis of Garcia \textit{et al.} \protect\cite{garc?a}.
}{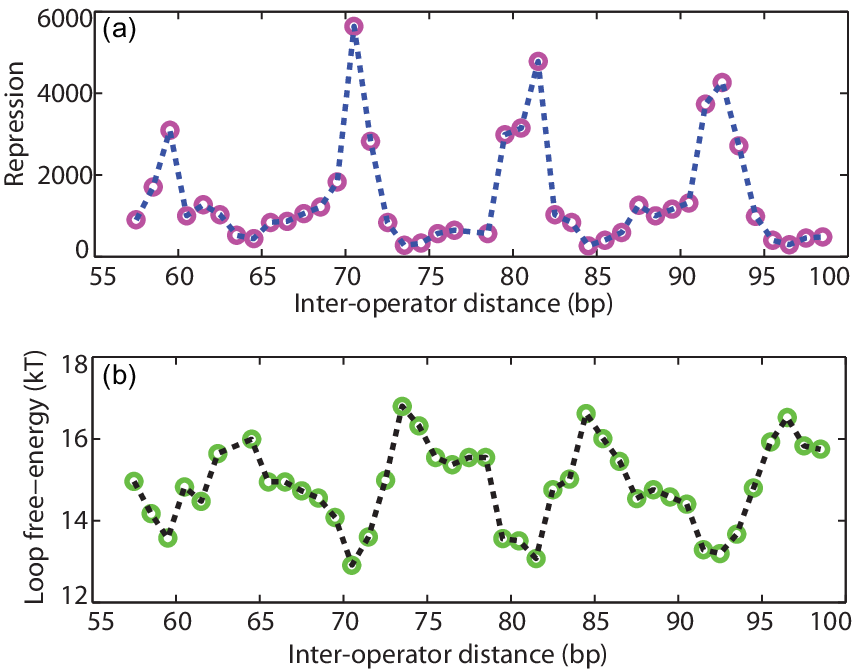}{3}

The interpretation of these results followed 
an analogy to the related 
process of DNA cyclization. Suppose that a regulatory
 protein binds
stereospecifically to the two operators, forcing them into close
physical proximity, with the intervening DNA forming a loop
(\fref{fone}). The equilibrium constant for this isomerization reaction
depends on the free energy change, which contains as a component the
elastic energy cost of forming the loop. The elastic energy, in turn,
contains terms reflecting bending and twisting deformation. For a
favorable value of the interoperator spacing, loop
formation involves only bending of the DNA. For spacing differing
slightly from 
the optimum, however, bringing the loop ends into the
relative orientation imposed by the protein complex requires an
additional rotation of one end about its axis. The extra elastic energy cost entailed by
this deformation reduces the equilibrium constant for loop formation relative
to the optimal spacing. But if the spacing is increased by a full
helical turn $\phelix$ (about 11 basepairs\pfoots\footnote{Although the canonical value of DNA pitch 
$\phelix$ is quoted as 10.5bp, this
value in fact depends on the temperature, solution conditions, superhelical
stress and so on. In this paper we will use the value 11bp as an approximation
to the actual period.}), then we once again have a twist-free
loop solution, a relatively low elastic energy cost, and hence a
relatively high level of gene regulation. Thus the hypothesis of loop
formation predicts a periodic modulation of the regulatory efficacy
with loop size, as observed. Mossing and Record put forward a
version of this theory shortly after the first
experimental results \cite{moss86a}.
 
\ifigureab{fone}{\textit{a.} Crystallographically derived structure of 
the lac repressor (LacI, \textit{dark gray)} bound to two operator segments of DNA
\textit{(black)} \protect\cite{lewi96a}. The \textit{light gray} curve 
represents a DNA conformation interpolating between the operator
segments, obtained in \protect\rref{bala99a}.
\textit{b.} Cartoon of the class 
of loops we will study. The DNA is
considered free in the region between the two
 exit points $s\pini$ and 
$s\pfin$. These exit  points are located at $\pm 10.5
\,$bp from the 
operator centers. Within the binding sites themselves, the protein may
induce kinks in the DNA, as shown.
\textit{c.} Definition of the initial and
 final tangent vectors $\pthat\pini$, $\pthat\pfin$, 
the separation vector $\pavec$, and the angle $\pthetaa$ characterizing our
idealized DNA--protein complex. $\pavec$ is the vector joining the two exit points, located at
arclengths $s\pini $ and $s\pfin$. The arclength separation between 
exit points is $\pspacing=
s\pfin-s\pini $.
 The vectors  $\pthat\pini$, $\pthat\pfin$, and $\pavec$ are
all assumed to be be coplanar, and the angle $\pthetaa$ from $\pavec$ 
to $-\pthat\pfin$ is equal to that from $-\pavec$ to $\pthat\pini$
(the ``planar, symmetric 
coupler'' idealization).
In the example shown, $\pthetaa>90^{\circ}$. Although the coupler is
planar, the loop itself will not in general be so, as illustrated 
here.
\textit{d.}~The $\star$-loop configuration corresponding to (c) (see
text \sref{s:rsls}). This loop is always planar.
}{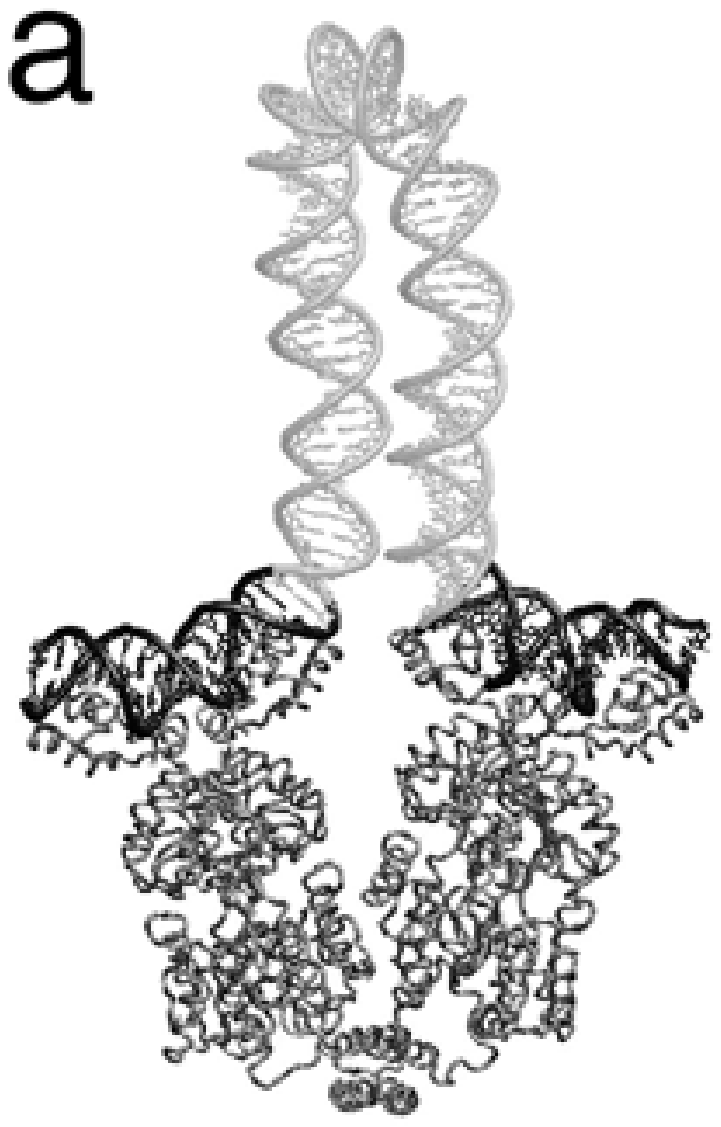}{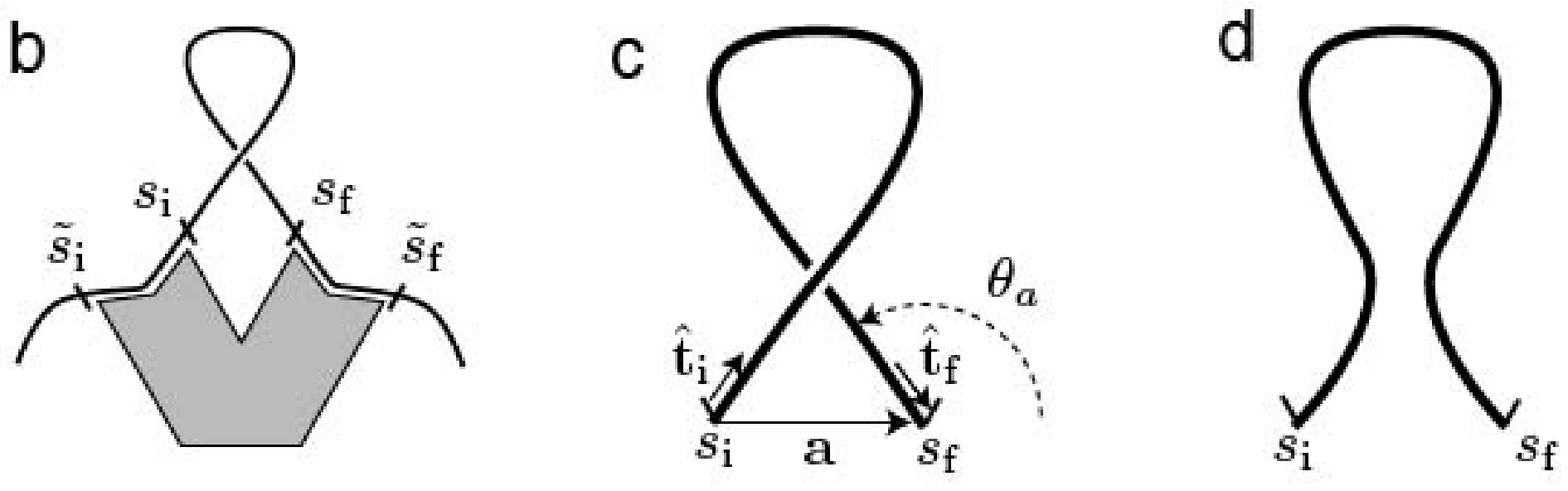}
{1.4}

Later, looped DNA complexes similar to those inferred 
by the above
argument were seen directly in electron microscopy (e.g. \cite{kram87a}) and
other modalities. More recently, single-molecule experiments have demonstrated DNA
looping \textit{in vitro}, and enabled the systematic study of the 
looping
reaction as a function of external parameters such as stretching force
\cite{finz95a,lia03a}. On the structural side, advances in x-ray
crystallography 
have yielded structures for the operator-protein
complex, for example in the \plac{}  operon system
\cite{lewi96a,bell01a,bell01b}. Starting
with that work, many authors have sought to determine the detailed form
of the loop using physical modeling (see
\sref{s:ippw}). A more ambitious goal would be to predict the looping free energy
function, which has recently been phenomenologically extracted from experimental studies of gene repression
{\it in vivo} (for example \cite{bint05b,saiz05a,saiz?a,garc?a}; see
\frefn{ftwo}b). This paper is intended as a step in that direction.

\subsection{Goals of this work\pnlabel{s:gw}}
Our goal in this paper is to introduce one important physical aspect
of looping, relevant 
both {\it in vivo} and in single-molecule assays.
This is the presence of significant torsional stress (supercoiling) in 
the region of DNA outside the loop-forming tract. Certainly everyday
experience teaches that external torque can predispose an elastic rod
(such as a garden hose) to form a loop.

A simple estimate shows that this external stress can significantly alter the 
equilibria between the unlooped state and various alternative looped
states. As we will review later, bacteria maintain their
chromosomal DNA in a negatively supercoiled (undertwisted) state, with 
a local torsional stress $\ptorq$ 
of about $-4\,$pN$\,$nm (see \sref{s:es}). Formation of a loop can
relax the external DNA by an angle on the order of $\pm\pi$ radians,
allowing the external torsional stress to do work
$\approx\pm\pi\ptorq\approx\pm3\,\kbt$ on the looping complex. This energy scale is
comparable to the looping energies inferred from data
(\frefn{ftwo}b), so we
must account for it. Indeed, previous authors have already
documented a large effect of supercoiling on a related process, the
juxtaposition of sequentially distant points on a long circular DNA
molecule \cite{klen95a,volo96a}. We wish to study similar effects in a simple
way, in the context of DNA looping. Specifically, we will calculate,
in a simplified model, the quasiperiodic dependence of the looping free
energy (\frefn{ftwo}b) on the interoperator spacing $\pspacing$.

We also give a procedure to find, in an idealized physical
model, the equilibrium shapes and energies of an elastic rod under the
sort of end constraints appropriate to DNA loop formation by a protein
complex. Our method uses the explicit analytic solutions to the
elastic-rod equations, and hence enjoys significant computational advantages
over gradient-descent algorithms.

Our results show that indeed external torque affects looping
equilibria, and can  change which of multiple looped states is
most favorable. Specifically, the shape of the looping free energy
curve reflects exchanges of stability as $\pspacing$ increases; the critical
values of $\pspacing $ for these exchanges (local maxima of \frefn{ftwo}b)
depend on the external torsional stress. These results can be tested, for example {\it in vivo} by studying 
bacteria with varying levels of supercoiling density
(\cite{volobook}, section 2.II.D), or \textit{in vitro} by 
the methods of Lia \textit{et al.} \cite{lia03a}. The methods developed in this paper
may also be applicable to other 
systems where DNA loops are implicated \cite{half04a}.

The paper is organized as follows. \sref{s:pf} outlines some prior work
and sets out the many simplifications we introduce to keep the treatment
of external supercoiling relatively transparent. \sref{s:cs} gives more
details of our calculation strategy. \sref{s:cd} presents
the actual calculation, and \sref{s:r} discusses the results. Expert readers
wishing to see the key new elements of our approach will find them
in \srefs{s:wpcd}--\ref{s:fsltpls} and \ref{s:et:3Dc}.

The Appendix gives a glossary of symbols used in the text.

\section{Physical framework\pnlabel{s:pf}}
In the first subsection below we describe some of the physical
ingredients that enter into the problem of modeling DNA looping. It is 
not possible  to survey here the large literature on such models, but we will
mention some of the prior work incorporating these ingredients. Mainly we
discuss work on the \plac{}  system, but extensive work has also been done
on other regulatory systems, such as \textit{gal} (e.g. \cite{gean01a}) and
lambda (e.g. \cite{bell00a}), and on the binding of
nucleosomes to miniplasmids (e.g. \cite{tobi00a}).
 
\subsection{Ingredients of the problem and prior work\pnlabel{s:ippw}}
\subsubsection{Loop structure\pnlabel{s:od}}
The crystallographic work of Lewis and coauthors gave only the
structure of the regulatory protein complex bound to two short DNA
segments containing the operators. Following this work, several
authors used elastic models of DNA to propose structures for the complete
looped state (for
example, \cite{bala99a,tsod99a,tobi00a,bala04a,bala04b,doua05a,bala06a}).
(Earlier authors studied similar mathematical problems in other
contexts, for example, \cite{tobi94a,cole95a}). The basic premise of these works
is that the regulatory protein complex binds to two specific elements
on the DNA, with a fixed, specified orientation relative to it (the
``strong anchoring end condition'' \cite{tobi00a}). The DNA between the
two binding regions must accommodate to these constraints by adopting a
form different from the one it would otherwise have taken; finding this
form is a boundary-value problem in the elasticity of a slender body.
These works included varying levels of realism in their treatment of
the DNA elasticity: Some included bend anisotropy, sequence dependence,
and electrostatic effects. Some, however, neglected DNA twist stiffness
altogether, and so could not address the periodic phasing dependence
that is part of our main motivation.

Several authors have recognized that there may be alternate 
DNA binding
patterns, giving rise to multiple looping states (for example
\cite{gean01a,swig06b,saiz?a}). We discuss this further in \sref{s:Dpb}
below. 

In addition to purely elastic effects, it has long been recognized that
the conformation of DNA is critically affected by chain entropy. An
early calculation including these effects was Shimada and
Yamakawa's study of DNA cyclization, the formation of circular DNA 
from linear pieces; later work has extended and refined their 
results \cite{shim84a,zhan03a,spak06a}. Recent work on DNA looping has begun
to incorporate entropic corrections following a similar calculational
approach \cite{zhan06a,swig06b,zhan?}. Although these corrections can 
be significant, for short loops the strong anchoring
condition constrains the DNA so much that elastic effects dominate over
conformational entropy (at least for understanding the periodic 
phasing dependence that is our central concern).

Other calculations have acknowledged that the protein complex formed in
DNA looping may not be a rigid object; stresses transmitted from the
bent DNA may distort the protein, or even induce major conformational changes in
it \cite{vill04a,vill05a,zhan06a,swig06b}.

\subsubsection{External supercoiling\pnlabel{s:es}}
In the absence of external constraints and thermally-induced
deformation, DNA would be a double helix with helical pitch
$\phelix\approx 11\,\mathrm{bp}\approx3.7\,\nmunit$. We define a corresponding 
quantity $\pono=2\pi/\phelix$, the angular rate at which the two
strands orbit their common centerline.

Closed circular DNA isolated from bacteria generally shows negative
supercoiling \cite{volobook}. This supercoiling is  expressed 
as the fraction $\sigma$ by which the total linking number differs from the
value $\pLtot\pono/2\pi$ appropriate for a torsionally relaxed, circular
loop; thus bacteria have $\sigma<0$.
The value of $\sigma$ can vary with the life conditions (e.g.
temperature) of the cell; it can vary from cell to cell and with the
division cycle of a single cell; and even
within a single cell, at one moment, there may effectively be domains of 
different $\sigma$ \cite{volobook}. 

Moreover, the topological linking number is not simply related to the quantity of
interest to us, which is the torsional stress $\ptorq$. First, in the
bacterial cell
various DNA-binding proteins can effectively absorb some linking number, similarly 
to the role of histones in eukaryotes. This binding results in a
reduced effective value  $\sigma\eff$ (sometimes called the ``superhelical
stress'') in the range of $-2.5$\% to $-5$\%
\cite{volobook,blis87a,drli92a,mark95b}. (Interestingly the corresponding value
for eukaryotes is close to zero \cite{volobook}.)

Second, even $\sigma\eff$ partitions into two components, corresponding to twist and writhe. Only the twist
component, roughly one quarter of the total \cite{volo92a}, gives rise to torsional stress
$\ptorq$. We estimate $\ptorq$ using Hooke's law, $\ptorq=\pktw
\Delta\omega$, where $\pktw\approx\xitbulkval \,\nmunit\,\kbt$ is the twist
stiffness of DNA under zero tension and
$\Delta\omega=(\frac14\sigma\eff)\pono\approx-0.017/\nmunit$ from the
above estimates. Thus $\ptorq\approx-1\kbt$, within the wide uncertainties
implied by the preceding paragraphs. In particular, the dispersion in
$\ptorq$ values implies that the observed repression curve
(\frefn{ftwo}a) will be an average over a distribution of $\ptorq$
values.

None of the prior work mentioned in \sref{s:od} introduced external torsional stress
(supercoiling) quantitatively; that is the goal of the present work.
This neglect is justified when studying loop formation in open (linear)
DNA segments. Even in the context of a circular, supercoiled DNA, the
strong anchoring condition implies that the interior of the loop is
unaffected by external torsional stress (if we neglect possible
stress-dependent deformation of the protein). Hence for any given
looped state, the geometric shape of the loop does not depend on this
stress. However, supercoiling does alter the equilibrium among the
different looped states, and between them and the unlooped state, and
hence it will affect curves such as those in \fref{ftwo}.

Swigon and coauthors do discuss the role of supercoiling qualitatively
\cite{swig06b}. As mentioned earlier, Vologodskii and coauthors also
studied its effect on site juxtaposition, in a large Brownian dynamics
simulation\cite{volo96a}. We seek a framework
for looping calculations in which such effects can be modeled, at least
approximately, without recourse to such large computations.

\subsection{Framework and idealizations used in this paper\pnlabel{s:fiup}}
We will make many simplifying assumptions in this paper, in order to
focus more clearly on effects of interest to us. Some of these
assumptions are already familiar from others' earlier work. Taken
together, these simplifications preclude detailed quantitative comparison with
experimental data like those in \fref{ftwo}. But 
the lessons we learn can be readily transferred to more detailed models.

\subsubsection{DNA} Although bending anisotropy, sequence dependence,  nonlinear DNA
elasticity \cite{wigg06a,wigg06b}, and perhaps even strand separation are likely to be
important to the quantitative understanding of loop formation, we
neglect them all. That is, we treat DNA as a continuous, inextensible, isotropic
elastic rod, with a linear relation between stress and the resulting
strain (the Bernoulli--Euler approximation \eref{e:BE}). We will also neglect
electrostatic effects, or more precisely, assume that they can be
effectively incorporated via effective values of the DNA bend stiffness
and the binding constants for the protein. The advantage
of these simplifications is that they will let us use the elegant closed-form solutions
to the elastic equilibrium equations (\sref{s:cd}).

Although the free DNA is assumed to be long, and so has significant
configurational entropy, as mentioned earlier we will neglect
fluctuations of the DNA inside the loop, and their entropy, because we
are interested in short loops. The ideas advanced in this article can be
applied to more elaborate calculations involving chain entropy effects.

We will assume that within the loop, we may neglect self-contact of the 
DNA. Thus we can only find the simplest one or two topoisomers in any
given situation, because higher topoisomers are
generally stabilized by self-avoidance. However, at least at moderate values of 
the external supercoiling, the higher topoisomers have very high
elastic energy and may indeed be neglected. 

Finally, we will assume that there are no other DNA-binding proteins in
the system that can bind to the loop region, altering its elasticity or
even imposing sharp bends on the DNA. In fact, at least one such
protein was present in the experiment of \fref{ftwo}, namely the heat    
             unstable nucleoid protein (HU). But
similar data can be obtained from mutant bacteria that are missing  particular
proteins (e.g. HU~\cite{beck05a}), and in any case {\it in vitro} 
assays can also be performed with no other proteins present.

\subsubsection{Protein} The repressor protein complex, like any protein, is 
flexible: It can
deform under stress, and in the case of LacI can even pop into 
an extended conformation. We
will neglect these effects,  treating the protein complex as a
rigid jig, or clamp, which we will call the ``coupler''\pfoots\footnote{This simplifying assumption may be 
more realistic for other complexes, such as the  lambda cI repressor,
which are thought to be more rigid than LacI.}. The geometry of the 
coupler is independent of the length of the DNA between the two 
operators.

\subsubsection{DNA--protein binding\pnlabel{s:Dpb}}
The LacI  protein complex is a tetramer
with two binding sites for DNA. Each binding site can bind strongly to 
specific operator sequences, or more weakly to generic DNA, or not at
all. Bintu and coauthors have argued that for LacI, {\it in vivo}, both sites
are nearly always bound to DNA; the strong binding to a few specific
sites competes with the weak binding to many generic sites
\cite{bint05b,garc?a}. We will 
instead simplify by assuming that the protein consists of two
halves, each
permanently bound to their operator sites. The looping reaction then
consists of these dimers finding and binding to each other, thus
imposing a fixed relative orientation on their bound operator 
DNA.\pfoots\footnote{Again, this idealization may be more literally appropriate
for other repressors, such as lambda cI.}

The specific binding of LacI at each site is thought to have a two-fold degeneracy
arising from the symmetry of each dimer: The operator DNA may be
reversed in direction and still 
bind equally well. This degeneracy
leads to four competing loop states \cite{gean01a,swig06b,saiz?a}. We will
neglect this complication, assuming that
only a single DNA orientation is allowed at each binding site. (The
binding orientations 
we choose produce what is often called the ``parallel
loop'' state \cite{swig06b}.) The equilibrium between distinct binding
orientations can be handled by the same methods as those used in the
present paper for the equilibrium between different looped states.

The geometry of the \plac{}  repressor complex is known to 
be chiral. Thus even in the absence of any
external torsional stress, the protein complex itself predisposes the
DNA to loop with a particular helical sense. One
contribution to this chirality comes from the fact that in the
cartoon of \frefn{fone}c, the arrows representing the required DNA
tangent vectors do not lie in the plane of the figure, but instead tilt
slightly into the page on their right sides, by an angle often called
$\beta$ \cite{swig06b}. We will neglect this effect,
assuming that the two boundary conditions correspond to tangent vectors
in the same plane as the separation vector between the detachment
points ($\beta=0$). We call this assumption the ``planar coupler'' condition. The 
methods of this paper can be extended to handle the case of nonzero $\beta$.
Note that even with a planar coupler, the DNA loop itself need not, and generally will not, be 
planar. Thus in the structure sketched in \frefn{fone}c, the DNA will
not in general contact itself in the middle of the loop.

Protein binding generally bends DNA, and in some cases untwists it
as well. Because we treat the protein as permanently bound to each
operator, we need not worry about these effects, as long as the
entrance points $\tilde s\pini$, $\tilde s\pfin$ (\frefn{fone}b), and
their corresponding tangent vectors, also lie in
the same plane as the one just mentioned. We thus add this 
requirement to our ``planar coupler'' condition.

There are two other sources of chirality (besides nonzero $\beta$ and protein-induced unwinding
mentioned above), which we do retain: First, as mentioned above, the
axial orientations of the two binding sites can induce a nonplanar
equilibrium shape for the DNA loop, even if the coupler obeys the
planar condition. Also, of course, any external supercoiling introduces
another chiral ingredient into the problem. We believe that 
these two effects are more important for the qualitative structure of \fref{ftwo}
than the twist angle $\beta$, and in any case they are the 
effects that we have chosen to study in this paper.

We also assume that the angle $\pthetaa$ shown in \frefn{fone}c equals
the corresponding angle on the left side (the ``symmetric planar
coupler;'' see also \cite{tobi94a}).  Our choice is motivated by approximate symmetries actually observed in
crystallographic data on protein--DNA complexes \cite{stei74a,lewi96a}.
The angle $\pthetaa$ may have a different
effective value in solution from the one observed in crystallographic
structures, so we will treat it as an unknown parameter in our analysis.
However, we take the separation $\pasca$ between the exit points to
have a fixed value $\pavalue \,\nmunit$ roughly equal to that seen in the
crystal structure \cite{lewi96a}.

\section{Calculation strategy\pnlabel{s:cs}}
\subsection{Mathematical representation\pnlabel{s:mr}}
We represent a thin elastic rod as a curve in space (the rod's
centerline), together with a set of orthonormal triads at each point on
the curve (the ``physical frame''). The third vector of each triad,
$\ehat _3(s)$, is chosen to coincide with the tangent to the curve at
arclength location $s$. We may choose $\ehat _1(s)$ to point from the
centerline toward the major groove of the DNA at position $s$, and $
\ehat_2(s)$ to complete the triad. Thus for relaxed 
DNA in the absence of thermal motion, as $s$ increases $\ehat _3(s)$ points in a
 constant direction while
$\ehat_{1,2}(s)$ rotate about it a  constant angular rate $\pono$ equal to $2\pi$ radians 
per helical turn. In general we say that a 
rod has zero excess twist if
the momentary rate of rotation of its physical frame has  3-component equal to
$\pono$.

For many purposes, it is convenient to replace the physical frame given
above by an ``untwisted frame'' obtained by rotating the physical frame
at each point about $\ehat _3(s)$ by the angle $-\pono s$. We will
denote the untwisted frame by $\dhat _i(s)$, and use it in the
calculations of \sref{s:et:3Dc}.

We represent the coupler mathematically as a condition specifying the
relative spatial locations and physical frames of the DNA as it exits
the two binding sites and enters the loop region (see \fref{fone}).
That is, stereospecific binding to the protein complex requires that
the location and orientation at positions $s\pini$ and $s\pfin$ are
related by a fixed element of the Euclidean group $E(3)$. In
particular, this relation is independent of the interoperator spacing
$\pspacing$.

We can express the same condition in the untwisted frame $\{\dhat_i\}$.
Now the relation between frames at $s\pini$ and $s\pfin$ does depend on
$\pspacing$, but in a simple way: Compared to the physical frame, the
required final orientation has an additional rotation about $\ehat_3(
s\pfin)$, by $-\pono\pspacing$.

For certain special values of $\pspacing$, we will be able to meet the 
coupler's boundary condition in a very simple way, with a loop that
stays in the plane determined by the coupler and has zero excess twist. These 
values take the form 
\begin{equation}\pnlabel{e:specialL}
    \pspacing=\pspacing_0+j\phelix,\end{equation}
where $j$ is any 
integer and $\pspacing_0$ is a constant depending on the coupler
geometry. For generic values of $\pspacing$, however, any loop must
either writhe out of the plane, or have twist density different from
$\pono$, or both.

\subsection{Why the problem is conceptually difficult\pnlabel{s:wpcd}}
Suppose that we are studying looping in a large, closed DNA molecule
($\pLtot=\,$thousands of basepairs), with a particular small separation
between the operator sites ($\pspacing=\,$dozens of basepairs). We divide all states into broad classes, or
``looping states'' (\fref{fthree}): those that are unlooped, and a
set of looped states.  The fraction of time spent in the unlooped state
determines the level of gene repression \cite{bint05b}, and is in turn 
determined by the relative free energies of the various states \cite{PhilBook}.

The transitions between looping states do not
change the total linking number of the full DNA molecule. Nevertheless,
there is a topological distinction between the classes of looped
states, which allows us to refer to them as ``topoisomers.''
To see this, imagine clipping out the looped regions of the looped
states in\frefn{fthree}{a}. The strong anchoring condition implies that we can
find a small reference arc $C$ such that each
such clipped region can be completed to a continuous closed loop by
gluing in the same piece $C$ (\fref{fthree}b). After this operation we can calculate the
linking numbers of the two resulting small closed loops, which will in
general differ by an integer.

\ifigure{fthree}{\textit{a.}~Equilibria between the unlooped state
$\pulst$ and various
looped states $\plst$, $\plst'$. We have omitted the regulatory proteins, indicating
 the
operator sites by tick marks. We wish to treat the system as a small
subsystem of interest (\textit{inside dashed line}), thermodynamically coupled to a large
reservoir (\textit{outside}).
Each looped state differs from the others by a $2\pi$
rotation of one strand of the DNA about its axis at its binding site. Thus, although the
total linking number is the same for every state shown, nevertheless
the set of looped states divides into classes labeled by an
integer. The two looped states shown have the same elastic energy 
inside the dashed lines, but quite different total free energy 
changes because of the torsional stress 
arising from supercoiling outside the dashed lines. 
\textit{b.}~One way to distinguish topological classes of looped states
is to choose a standard reference arc $C$ that completes each of the looped
configurations, then compute the linking 
numbers of the resulting closed loops.}{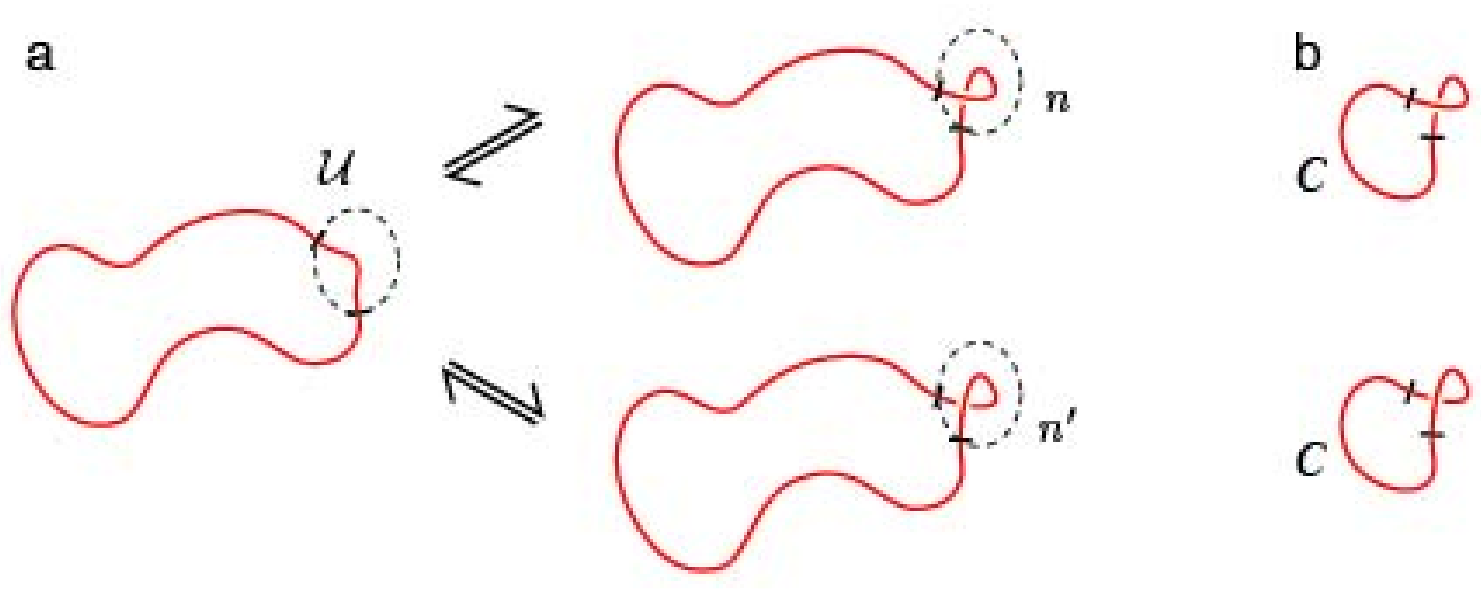}{2}

Clearly the equilibrium between the looped and the various unlooped
states will be affected by the initial degree of supercoiling in the
molecule. We would like to treat the region outside the looping
region as a ``reservoir'' and characterize it by a ``thermodynamic
force'' acting on the loop region. To see why this is not entirely
straightforward, we contrast to an easy problem: Suppose we have a small
box of air in contact with a large room  via a pinhole. For the
purposes of calculating the average number of gas molecules in the box 
$\langle N_1\rangle$, we can forget about the size and shape of the room, instead
characterizing it by a single number, the pressure. The rest of the
calculation is easy because there is a local, additive conservation law
relating $N_1$ to the number $N_2$ of molecules in the room, and
because the boundary between the two subsystems is fixed. In contrast, 
in our problem the linking number, although conserved, is not locally
defined, and the two operator sites are free to move in the unlooped
state.

\subsection{The $\star$-loop state\pnlabel{s:rsls}}
\subsubsection{Decomposition of the free energy change}
Our problem would be easier if we had only to investigate the
equilibrium between various looped states, not that between looped and 
unlooped states! After all, a direct transition between the states
$\plst$ and $\plst'$ in \frefn{fthree}a simply involves an axial
rotation by $2\pi$. In the limit where the total
DNA length $\pLtot \gg\pspacing$, the external torsional stress $\ptorq$ is
constant during this process, so the exterior region does work on the looping
 region given by $2\pi\ptorq$. Adding this work to the
change in elastic energy 
gives the total free energy change of the
transition.

To see how to extend the above remark to include loop formation,
we found it useful to introduce a fictitious
looping state, which we call the $\star$-state, and to divide the free 
energy change of looping into two pieces: That for the transition from 
unlooped to the $\star$-state, and that for a subsequent transition
to the desired physical looped state. 

The $\star$-state is characterized by a modified $\star$-coupler, 
identical to the actual coupler except for the axial orientation it 
imposes on the outgoing DNA, which is always chosen to admit a planar, 
untwisted loop regardless of $\pspacing$. One such loop is a 
non-selfcontacting solution to the elastic equilibrium equations; we call it the $\star$-loop
configuration (see \frefn{fone}d).

Thus we write  the free energy change for formation of looped state 
$\ploopn$ as
\begin{equation}\pnlabel{e:decomp}\Delta G_{{\rm open}\to{\rm 
    loop}\,\ploopn}=
\Delta G_{{\rm 
open}\to{\star}}+\Delta G_{{\star}\to{\rm loop}\,\ploopn}
\end{equation}
We wish to calculate each term on the right. In fact, the second term can be
evaluated by the same method as outlined in the first paragraph of this
subsection. We now turn to discuss the first term.

\subsubsection{$\star$-loop formation free energy\pnlabel{s:slfe}}
In the unlooped state,
the full circular DNA is freely fluctuating. It has a certain free energy, which we assume to be 
extensive (at least over the small length changes we are studying):
$G\puns(\pLtot,\sigma)=\pLtot\mu(\sigma)$, where the free energy density
$\mu$ depends on the fractional degree of supercoiling $\sigma$. We
imagine cutting the DNA, introducing a full extra unit of linking
number, and resealing it. Examining the resulting change of free energy yields
a formula for the external torsional stress $\ptorq$:
\begin{equation}\pnlabel{e:torq}
    \frac{\dd\mu}{\dd\sigma}=\pono\ptorq.\end{equation}

We now turn to loop formation. The $\star$-loop is planar and untwisted. 
Thus its formation not only reduces the length of the remaining free
exterior region from $\pLtot$ to $\pLtot-\pspacing$; it also expels some 
linking number from the looped region, changing $\sigma$ to $\sigma'=\sigma
\pLtot/(\pLtot-\pspacing)\approx\sigma(1+(\pspacing/\pLtot))$.
Neglecting higher orders in $\pspacing/\pLtot$, the
corresponding change of free energy is thus
\begin{eqnarray}
\Delta G_{\rm 
bind}+G\puns(\pLtot-\pspacing,\sigma')-G\puns(\pLtot,\sigma)&\approx&
\Delta G_{\rm bind}+(\pLtot-\pspacing)\mu\bigl(\sigma+\frac{\pspacing\sigma}{\pLtot}\bigr)-G_{\rm
free}(\pLtot,\sigma)\nonumber\\
&\approx&\Delta G_{\rm bind}+\pspacing(-\mu(\sigma)+
\sigma\pono\ptorq)\nonumber \\
&\approx&\Delta G_{\rm bind}-\pspacing\mu(0)\pnlabel{e:getGfree}
\end{eqnarray}
Here $\Delta G_{\rm bind}$ is the binding free energy for assembly 
of the protein complex\pfoots\footnote{Recall that we are assuming that the
DNA is permanently bound to the protein. 
More generally we need to
account for the fact that protein--DNA binding generally unwinds the
DNA; for example in {\it lac} the unwinding is nearly one radian. The
work done by external torsional stress against this rotation
effectively modifies the binding constant relative to the value
appropriate for isolated operator fragments.}.
The total  free energy change $\Delta G_{{\rm open}\to{\star}}$ is the 
quantity in \eref{e:getGfree} plus the elastic energy $\pE\st$ of the
$\star$-loop (recall that we neglect the conformational entropy of the looped
regions).

The free energy $G\puns$ of supercoiled DNA has been investigated both
theoretically and experimentally \cite{mark95b}.
Rather than attempting to evaluate it explicitly, we now
just observe that, \eref{e:getGfree} is a fixed, linear
function of $\pspacing$; it does not contribute to the quasiperiodic modulation seen in
\fref{ftwo}, and it does not depend on which looped state
 $\ploopn$ we 
will eventually form. We
can drop these parts of the free energy 
change if we understand that 
our result will be correct only modulo the addition of some linear
function of $\pspacing$ to our calculated free energy change of looping.
This limitation does not impair our ability to predict the periodic 
modulation of the free energy change, nor to find nonlinear behavior
such as a dip in its envelope at a particular value of $\pspacing$, nor
to investigate the equilibrium between competing topoisomers (various 
$\ploopn$), nor to explore the $\sigma$ dependence of the looping free 
energy.

Again, henceforth we will drop the contributions to the looping free
energy given by \eref{e:getGfree}, or in other words we take $\Delta
G_{{\rm open}\to\star}=\pE_{\star}$ in \eref{e:decomp}.

\subsection{From $\star$-loop to physical looped states\pnlabel{s:fsltpls}}
For the special values $\pspacing=\pspacing_0+j\phelix$ mentioned in \sref{s:mr}, the
$\star$-state coincides with one of the physical looped states. For other
values of $\pspacing$, the $\star$-state is a useful intermediate,
because as we have seen its formation has a simple effect on the DNA outside the
looping region, and so does the passage from it to the actual looped
states.

Our procedure, then, is the following. We begin by choosing starting guesses for
the unknown parameter $\pspacing_0$ describing the periodically spaced 
special values of $\pspacing$, and the poorly known
parameter $\pthetaa$. We set reasonable values for the remaining
parameters $\ptorq\approx-1\kbt$, $\pasca\approx\pavalue \,\nmunit$ and for the
elastic constants of DNA.

We then step through the various values of $\pspacing$ in the range of
interest. For each $\pspacing$, we construct the $\star$-loop
(\sref{s:pl} below) as the planar, non-selfintersecting loop that
meets all the boundary conditions imposed by the coupler except for
axial orientation, and solves the elastic equilibrium equations. We call its elastic energy $\pE\st$.

If $\pspacing$ is one of the special values, then the $\star$-loop is
one of the possible physical looped states. Whether or not this is
true, we next perturb the
$\star$-loop through a family of nonplanar solutions to the elastic rod
equilibrium equations, maintaining always the same relative position and
tangents at the ends (\sref{s:ss} below). Each solution in this family has a final
orientation differing from the $\star$-loop by
an axial rotation.  
The corresponding rotation angle $\ppsi$ is a real
number (not ambiguous modulo 
$2\pi$). Each time $\ppsi$ passes through 
a value 
\begin{equation}\pnlabel{e:psic}
\ppsi_\ploopn =(\pspacing-\pspacing_0)\pono+2\pi \ploopn \quad\mbox{ for an
 integer $\ploopn $,}
\end{equation}
we get a physical looped
state. The angle $\ppsi_\ploopn $ may be either positive or negative
(or zero if $\pspacing$ takes one of the special values). 
We compute the elastic energy of this loop and call it $\pE_\ploopn $. For each 
physical loop found, we 
correct the energy to $\pE'_\ploopn =\pE_\ploopn -\ppsi_\ploopn \ptorq$ to account for the external
torsional stress, obtaining $\Delta
G_{\star\to{\rm loop\,}\ploopn}=\pE_n -\ppsi_\ploopn \ptorq-\pE\st$.

The quantity $\pE\st$ cancels when we compute the total free energy 
change (\eref{e:decomp}); as described in \sref{s:slfe}, we  also
drop the linear contribution \eref{e:getGfree}. 
Thus 
\begin{equation}\pnlabel{e:DG}
\Delta G_{\mathrm{open}\to\mathrm{loop\ }\ploopn }=\pE'_\ploopn=\pE_\ploopn -
\ppsi_\ploopn \ptorq
,\end{equation}
modulo the addition of a fixed, linear function of $\pspacing$.

\eref{e:DG} embodies one of the main points of this paper. We can
understand it physically by thinking about \frefn{fthree}a: The two
looping states $\ploopn$ and $\ploopn'$ have the same elastic energy,
but one is favored and the other disfavored by external torsional
stress. The correction term in \eref{e:DG} incorporates this effect.

In general, we will only obtain one or two solutions in this way for
each value of $\pspacing$; as mentioned earlier, higher topoisomers are
stabilized by self-contact, which our model neglects. We now plot each
energy value $\pE'_\ploopn $ versus its $\pspacing$. Taking the smallest $\pE'_\ploopn $
for each $\pspacing$ gives a graph (\fref{fig:twisttheory} below). Finally, we repeat the whole procedure with different
values of the parameters $\pspacing_0$ and $\pthetaa$, and seek values 
that are reasonable and that roughly reproduce experimental data like those
in \fref{ftwo}.

\section{Calculation details\pnlabel{s:cd}}
We now give details of the calculation outlined in the 
previous
section.
\subsection{Two dimensional warmup problem\pnlabel{s:plane}}
\ifigure{fig:loopfig}{The geometry of our 2D exercise. The contour length of the 
DNA in the loop is $L$. The size of the protein complex is represented by $a$; 
its geometry is summarized by  the parameter $\theta_{a}$. $a$ and $\theta_{a}$
determine the boundary 
 conditions for the boundary value problem for the
 geometry of the loop. The 
 element at arclength $s$ from the center exerts a force $\pnvec{F}$ on the element at
 $s + \dd s$. Due to the assumed symmetry, $\pnvec{F}$ points along the negative
 $z$-axis as
 shown. $\pnvec{F}$ is also the
 sideways force exerted on the ends of the rod by 
 the protein complex.}{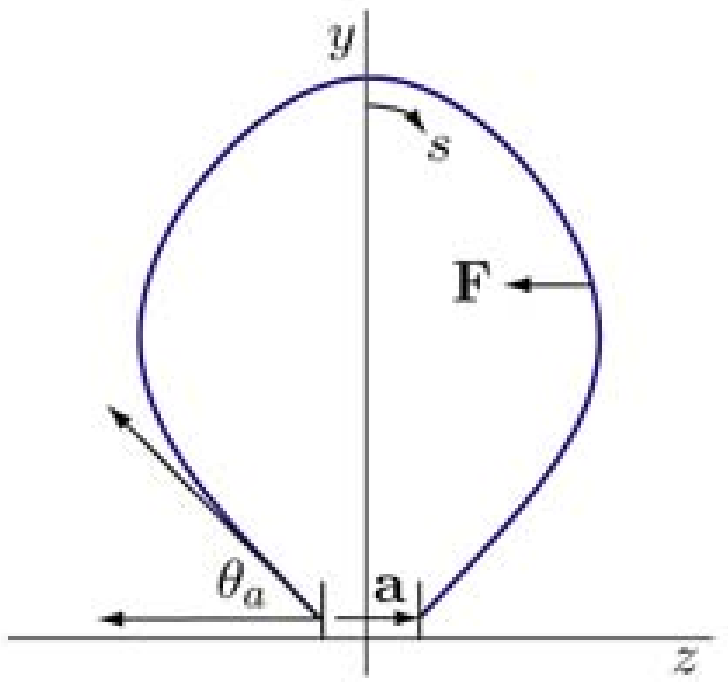}{2}
As a warmup problem, we consider the analogous elasticity problem in
two dimensions. That is, we find the profile $z(s)$ and $y(s)$ of a planar elastic 
loop (\fref{fig:loopfig}) where $s$ denotes the arc-length along the 
loop with the origin $s=0$ placed midway along the contour. Our equations will be simple
because twist elasticity plays no role in two dimensions. 
  
The boundary
 value problem for the loop can be stated as 
\begin{equation} \pnlabel{eq:firsteq}
 \pkb \theta'' + F \sin\theta  = 0, \qquad \theta(0) = 0, 
 \quad \theta(\lfr{L}{2}) = \pi + \theta_{a},
\end{equation}
where $\pkb $ is the bending modulus of the elastic rod, $\theta(s)$ 
is the angle from the positive $z$-axis to the
 tangent and $\pnvec{F}$ is an unknown force acting along the $z$-axis and 
exerted by the end-clamp on the rod. Primes denote differentiation with respect
to the arclength $s$. 
 
The solution to the second order differential equation above is well known 
\cite{niz99a} and can be written as
\begin{equation}
 \theta(s) = 2\am(\frac{s}{\lambda k}|k),
\end{equation}
where $\am(x|k)$ is the elliptic function of the first kind with modulus $k$.
$\lambda = \sqrt{\frac{\pkb}{F}}$ and $k$ 
are independent parameters, to be determined from the boundary data. Thus 
\begin{eqnarray}
 \cos\theta(s) = \frac{\dd z}{\dd s} & =&  1 - 2\sn^{2}(\frac{s}{\lambda k}|k), \\
 \sin\theta(s) = \frac{\dd y}{\dd s} & =&  2\sn(\frac{s}{\lambda k}|k)
                                      \cn(\frac{s}{\lambda k}|k).
\end{eqnarray} 
We can integrate these equations and obtain the following solution for $z(s)$ 
and $y(s)$, which are the coordinates of the material point denoted by $s$ 
on the rod:
\begin{eqnarray}
 z(s) & =& s - 2\int_{0}^{s} \sn^{2}(\frac{\alpha}{\lambda k}|k)\dd\alpha, \\
 y(s) & =& \int_{0}^{s} 2\sn(\frac{\alpha}{\lambda k}|k) 
                         \cn(\frac{\alpha}{\lambda k}|k) \dd\alpha 
        = \frac{2 \lambda}{k}(1- \dn(\frac{s}{\lambda k}|k)).
\end{eqnarray}
\fref{fig:loopfig} shows a typical solution from this family.

The two constants $\lambda$ and $k$ can be determined in terms of the 
given $\pasca$ and $\pthetaa$ by imposing the 
boundary conditions, leading to the following two equations:  
\begin{eqnarray}
 \pi + \theta_{a} & =&\theta(\frac{\pspacing}{2})= 2\am(\frac{L}{2\lambda k}|k), \\
 \frac{a}{2} & =& \frac{L}{2}  - 2\int_{0}^{L/2} \sn^{2}(\frac{\alpha}{\lambda k}|k)\dd\alpha.
\end{eqnarray}
We denote $y_{p} \equiv \frac{L}{2\lambda k}$ and eliminate $\lambda$ in favor of 
$y_{p}$, obtaining 
\begin{eqnarray}
 \pi + \theta_{a} & =& 2\am(y_{p}|k), \\
 y_{p}(1 - \frac{a}{L}) 
& =& 2\int_{0}^{y_{p}} \sn^{2}(\beta|k)\dd\beta 
                         = \frac{2}{k^{2}}(y_{p}-E(y_{p}|k)), 
\end{eqnarray}
where $E(y_{p}|k) = \int_{0}^{y_{p}} \dn^{2}(x|k)\dd x$ is the incomplete elliptic
integral of the second kind with modulus $k$. 
Once we solve numerically for $y_{p}$ 
and $k$ from these equations we can obtain the unknown force $F$ as 
\begin{equation}\pnlabel{e:Flamb} 
F = \frac{\pkb }{\lambda^{2}} = \frac{4\pkb y_{p}^{2}k^{2}}{L^{2}}.  
\end{equation}
Also, the bending moment $M$ applied by the protein at $s = \frac{L}{2}$ is given by 
\begin{equation}
 M = \pkb \theta'(\lfr{L}{2}) = 
     4\pkb \frac{y_{p}}{L}\sqrt{1 - k^{2}\cos^{2}\frac{\theta_{a}}{2}}.
\end{equation}

Finally we calculate the elastic energy stored in the loop.
\begin{eqnarray}
 \pE_{\rm elas}[\theta(s)] & =& \int_{-L/2}^{L/2}\Big(\frac{\pkb }{2}
 \theta'^{2}(s) -F\cos\theta(s)\Big)\dd s  \nonumber \\
 & =& F \int_{-L/2}^{L/2}\Big(\frac{2}{k^{2}}\dn^{2}(\frac{s}{\lambda k}|k) 
      + 2\sn^{2}(\frac{s}{\lambda k}|k) - 1\Big) \dd s \nonumber \\
 & =& F \int_{-L/2}^{L/2}\Big(\frac{2}{k^{2}} - 1\Big)  
   = F
L(\frac{2}{k^{2}}-1). \pnlabel{eq:planenerg} 
\end{eqnarray}
Similar formulas have appeared in earlier work (e.g. \cite{cole95a}).

It is well known that the equations describing the shape of a bent rod are
similar to the equations of motion of a pendulum and that this analogy can be
utilized to obtain rod shapes corresponding to different regimes 
\cite{niz99a}. 
In the above we looked at the solution corresponding to the revolving orbits
of the pendulum. We can also have solutions corresponding to oscillating 
orbits of the pendulum. In that case the solution is given by 
\begin{equation}
 \cos\theta(s) =1-2k^{2}\sn^{2}(\frac{s}{\lambda}|k).
\end{equation}
Corresponding to this solution we find that
\begin{equation}
 F = \frac{4\pkb y_{p}^{2}}{L^{2}},\qquad M = \frac{4y_{p}\pkb }{L}    
 \sqrt{k^{2} - \cos^{2}\frac{\theta_{a}}{2}}, \qquad  
 \pE_{\rm elas}[\theta(s)] = FL (2k^{2}-1). 
\end{equation}

\subsection{Elasticity theory: 3D calculation\pnlabel{s:et:3Dc}}
\subsubsection{$\star$-loop\pnlabel{s:pl}}
The $\star$-loop is by definition a planar, untwisted solution of the
elastic equilibrium problem with given separation and tangent vectors
at the ends. As such its centerline coincides with the 2D solution found in
\sref{s:plane} above. Its physical frame has $\ehat_3$ and $\ehat_1$
always in the $yz$ plane, and $\ehat_2$ along $\uvec x$.

\subsubsection{Rod equilibrium}
We now repeat our exercise for a uniform,
inextensible, isotropic elastic rod with twist 
stiffness, not necessarily confined to a plane. We again idealize the 
protein complex forming the loop as a rigid object attaching to two
specified points (representing specific binding sites) on the rod.
We assume that the length of the vector connecting one binding site to the other is given,
and we call it $a
 = |\mathbf{a}|$. The orientation of the physical 
frame attached at one site relative to the one attached at the other site, as well as
the orientation of $\mathbf{a}$ relative to either of those frames, are 
also assumed to be given.
The derivatives of the
 untwisted frame vectors $\{\dhat_i(s)\}$ as the arc-length $s$
changes contain 
information
about the local curvature and torsion of the rod:
\begin{equation}
\dhat _{i}' = \bkap \times 
 \dhat _{i}, \qquad \mbox{for} \quad i=1,2,3.
\end{equation}
We define $\kappa_{1,2,3}$ as the components of the curvature vector 
$\bkap (s)$ expanded in the frame $\{\dhat_i\}$. 

The moment (or torque) $\mathbf{M}(s)$ at any point on the rod is
assumed to have the Bernoulli--Euler form
\begin{equation}\pnlabel{e:BE}
 \mathbf{M} = \pkb \kappa_{1}\dhat _{1} + \pkb \kappa_{2}\dhat _{2}
 + \pktw \kappa_{3}\dhat _{3},
\end{equation}
where $\pkb
 $ and $\pktw $ are the bending and twisting moduli of the 
elastic rod.\pfoots\footnote{Note that the external torque, $\ptorq$, need not equal 
$M_3$ at the points $s\pini,s\pfin$, because the coupler can exert  torque on the DNA.}
Equivalently we can specify the persistence lengths $\pxib  = \pkb /\kbt $
and $\pxitw
  = \pktw /\kbt $, where $\kbt $ is the thermal energy scale.
The equilibrium equation for the rod is then simply given by
force balance, $\mathbf{F}' = \mathbf{
0}$, and by torque balance: 
\begin{equation}
 \mathbf{M}' + \dhat _{3}\times\mathbf{F} = \mathbf{0}.
\end{equation}
As in \fref{fig:loopfig}, $\mathbf{F}(s)$ is the force each element
exerts on the next; it is also the force applied by the protein on the DNA.  

Following Nizette and Goriely \cite{niz99a}, we will assume that the 
laboratory coordinate frame is chosen in such a way that the constant internal 
force $\mathbf{F}$ is aligned with the laboratory $z$-axis:
$\mathbf{F} = F\uvec{z}$.
We write the position vector $\mathbf{P}(s)$ of any point on the loop 
as $[X(s),Y(s),Z(s)]$ or in cylindrical coordinates as $[R(s),\Phi(s)
,Z(s)]$: 
\begin{eqnarray}\pnlabel{eq:posvecdef}
 \mathbf{P}(s) & =& X(s)\uvec{x
} + Y(s)\uvec{y} 
 + Z(s)\uvec{z} \nonumber \\ 
 & =& R(s)\cos\Phi(s)\uvec{x} + 
R(s)\sin\Phi(s)\uvec{y}
 + Z(s)\uvec{z} .
\end{eqnarray}  
Because we assume that our protein complex obeys the symmetric coupler condition 
(\sref{s:Dpb}), we will be interested in loops that are also symmetric 
in a way that generalizes \fref{fig:loopfig}.  Specifically, we will
find suitable equilibrium solutions satisfying our boundary conditions,
which also obey
\begin{equation} \pnlabel{eq:thetsug}
 X(s) = -X(-s), \qquad Y(s) = Y(-s), \qquad Z(s) =
 -Z(-s).
\end{equation}   
\erefs{eq:thetsug} reduce to our planar form when $X(
s) \equiv 0$. It may appear to treat the $X$ and $Y$ directions 
asymmetrically, but what is meant is that the solutions of interest can be 
brought to this form by translation and rotation about $\uvec{z}$. Thus for example,
if the loop is planar then we agree to place it in the $yz$-plane, with the
center point $s=0$ at the origin.
\erefs{eq:thetsug} suggest that $\Phi(s)$ will take the form $\Phi(s) 
= \frac{\pi}{2} - \alpha(s)$ with $\alpha(-s) = -\alpha(s)$, and indeed
our 
solutions will have this property. 

\subsubsection{Boundary conditions}
We are 
now in a position to formulate the boundary condition describing the 
relative position of the ends of the loop.
Our first condition again imposes a separation 
of length $a$:
\begin{equation} \pnlabel{eq:bc1}
 \|\mathbf{P}(\lfr{L}{2}) - \mathbf{P}(-\lfr{L}{2})\|^{2} = a^{2},
\end{equation}
where again $L$ is the loop contour length. Taking account of the symmetry of 
the loop, this statement amounts to
\begin{equation} \pnlabel{eq:bc2}
 R^{2}(\lfr{L}{2})\cos^{2}\Phi(\lfr{L}{2
}) + Z^{2}(\lfr{L}{2}) 
 = \frac{a^{2}}{4}.
\end{equation}
Notice that our choice of loop orientation implies that the end-to-end vector
of the loop $\mathbf{a}$ lies in the $xz$-plane, though not in general 
along the $z$-axis as in \fref{fig:loopfig}. 

The tangent vector at any point on the rod is given by 
$\hat{\mathbf{t}}(s) = \frac{\mathbf{P}'(s)}{\|\mathbf{P}'(s)\|}$. Explicitly, 
\begin{equation}
 \mathbf{P}'(s) = 
 (R'(s)\cos\Phi(s) - R(s)\Phi'(s)\sin\Phi(s))\uvec{x
}
 + (R'(s)\sin\Phi(s) + R(s)\Phi'(s)\cos\Phi(s))\uvec{y}
 + Z'(s)\uvec{z}
\end{equation}
which can be rewritten as
\begin{equation}
 \hat{\mathbf{t}}(s) = \mathbf{P}'(s) 
 = T(s)\cos\phi(s)\uvec{x} + T(s)\sin\phi(s)\uvec{y}  
   + \cos\theta(s)\uvec{z}
\end{equation}
where we define 
$\theta(s)$, $\phi(s)$ and $T(s)$ through
\begin{eqnarray}
 \cos\theta(s) & =& Z'(s), \\ 
 \tan\phi(s) & =& \tan\left(\Phi(s) +
 \tan^{-1}\frac{R(s)\Phi'(s)}{R'(s)}\right),\pnlabel{e:tanf} \\
 T^{2}(s) & =& R'^{
2}(s) + R^{2}(s)\Phi'^{2}(s) = 1 - Z'^{2}(s).
\end{eqnarray} 
The last equation 
reflects the inextensibility of the rod. 

Our planar coupler requires that the 
vectors $\mathbf{a}$, 
$\mathbf{P}'(-\frac{L}{2})$ and $\mathbf{P}'(\frac{L}{2})
$ all lie in a 
common plane, even if the full loop is not planar. 
Accordingly,
 we will seek solutions satisfying a second boundary condition:
\begin{equation}
 \pnlabel{eq:bc3}
 \mathbf{a}\cdot(\mathbf{P}'(-\lfr{L}{2})\times\mathbf{P}'(\lfr{L}{2})) = 0.
\end{equation}
Using the assumed symmetry of the shape of the loop, we see that this 
boundary condition can be satisfied if either  
\begin{equation} \pnlabel{eq:plane1} 
 \frac{a_{x}}{a_{z}} = 
 \frac{R(\frac{L}{2})\cos\Phi
(\frac{L}{2})}{Z(\frac{L}{2})} =  
 \frac{R'(\frac{L}{2})\cos\Phi(\frac{L}{2}) 
 - R(\frac{L}{2})\Phi'(\frac{L}{2})\sin\Phi(\frac{L}{2})}{Z'(\frac{L}{2})},
\end{equation}
or 
\begin{equation} \pnlabel{eq:plane2}
 R'(\lfr{L}{2})\sin\Phi(\lfr{L}{2}) 
 + R(\lfr{L}{2})\Phi'(\lfr{L}{2})\cos\Phi(\lfr{L}{2}) = 0.
\end{equation}

Substituting \eref{eq:plane2} into \eref{e:tanf}
shows that the second of these
conditions would imply $\phi(\pm\frac{L}{2}) =0$. This in turn would imply that
the end tangents $\mathbf{P}'(\pm\frac{L}{2})$ are parallel and lie on the $xz$-plane.
This violates the assumed geometry of the coupler depicted in 
\fref{fone}:
The end tangents need not be parallel. So in this section we
pursue only the condition represented by \eref{eq:plane1}, which allows 
for loops with the desired coupler geometry. 

The third boundary condition that we need to satisfy involves the angle at 
which the DNA exits the protein complex. Generalizing \eref{eq:firsteq}, 
we will require that  
\begin{equation} \pnlabel{eq:bc7}
 \mathbf{a}\cdot\mathbf
{P}'(\lfr{L}{2}) = a\cos\theta_{a}.
\end{equation}
Together \erefs{eq:bc1}, \ref{eq:plane1} and \ref{eq:bc7} can be recast
as the boundary conditions
\begin{eqnarray}
 \frac{Z'(\frac{L}{2})}{Z(\frac{L}{2})} & =&
 \frac{2\cos\theta_{a}}{a},\pnlabel{e:bca} \\
 R^{2}(\lfr{L}{2})\cos^{2}\Phi(\lfr{L}{2}) + Z^{2}(\lfr{L}{2}) 
 & =& \frac{a^{2}}{4}, \\
 \frac{R'(\frac{L}{2})}{R(\frac{L}{2})} - \Phi'(\lfr{L}{2})
 \tan\Phi(\lfr{L}{2}) & =& \frac{2\cos\theta_{a}}{a}.\pnlabel{e:bcc}
\end{eqnarray}

We will solve these equations starting from the most general solution to 
the differential equations for the equilibrium of a bent and twisted rod.
The solution yields analytical expressions for $R(s)$, $Z(s)$ and 
$\Phi(s)$, which will be substituted in the above to obtain algebraic equations. 
Nizette and Goriely \cite{niz99a} give the solution in terms of four
parameters $\pz_{1,2,3}$ and $\lambda$ as
\begin{eqnarray}
 Z(s) & =& \pz_{3}s - \lambda (\pz_{3}-\pz_{1})E(\frac{s}{\lambda}|k), \\
 R^{2}(s) & =&2\lambda^{2}(\tilde{h} - Z'(s)), 
\\ 
 \Phi(s) & =& \frac{\lambda}{2\pkb }\Big(M_{z}\frac{s}{\lambda} 
 + \frac{M_{3}-M_{z}\tilde{h}}{\tilde{h}-\pz_{1}}
 \Pi(\frac{s}{\lambda}|\tilde{n},k)\Big) 
- \frac{\pi}{2}, \end{eqnarray}
where $\Pi$ is the elliptic integral of the third kind and 
\begin{eqnarray}
M_{3} & =& \frac{\pkb}{\sqrt{2}\lambda}
  \left(\sqrt{(1+\pz_{1})(1+\pz_{2})(\pz_{3}+1)} +
  \sqrt{(1-\pz_{1})(1-\pz_{2})(\pz_{3}-1
)}\right), \\ 
M_{z}  & =& \frac{\pkb}{\sqrt{2}\lambda}
  \left(\sqrt{(1+\pz_{1}
)(1+\pz_{2})(\pz_{3}+1)} -
  \sqrt{(1-\pz_{1})(1-\pz_{2})(\pz_{3}-1)}\right), \\ 
 \tilde{h} & =& \frac{1}{2}\Big[\pz_{1} + \pz_{2} + \pz_{3} - \pz_{1}\pz_{2}\pz_{3}
 + \sqrt{(1-\pz_{1}^{2})(1-\pz_{2}^{2})(\pz_{3}^{2}-1)}\Big], \\
 \tilde{n
} & =& \frac{\pz_{2} - \pz_{1}}{\tilde{h}-\pz_{1}}, \qquad 
 k^{2} = \frac{\pz_{
2}-\pz_{1}}{\pz_{3} - \pz_{1}} \qquad 
 n = \frac{\pz_{2}-\pz_{1}}{h-\pz_{1}}.
\end{eqnarray} 
The quantities $M_3$ and $M_z$ turn out to be the components of the
moment vector along $\dhat_3$ and $\uvec z$, respectively \cite{niz99a}.

We must now fix the parameters $\pz_{1,2,3}$ and $\lambda$ by imposing 
boundary conditions. In addition to \erefs{e:bca}--\ref{e:bcc},
we also need an expression
for how the  frames at each end of the rod are oriented with respect
to each other. To formulate such an expression, note that for any choice of
$a$ and $\theta_{a}$ there will be one non-selfintersecting loop 
solution with zero excess twist --- the $\star$-loop. Taking this as a reference configuration,
any other equilibrium solution with the same $a$ and $\theta_{a}$ and the 
same initial $\dhat _{1}(-\frac{L}{2}) = \dhat _{1,{\rm ref}}(-\frac{L}{2})$
will have a final orientation $\dhat _{1}(+\frac{L}{2})$ differing from
$\dhat_{1,{\rm ref}}(+\frac{L}{2})$ by a rotation about 
$\hat{\mathbf{t}}(+\frac{L}{2})$. We need to find 
the corresponding 
rotation angle $\ppsi$, then impose the condition \eref{e:psic}.

The angle $\ppsi$,
divided by $2\pi$, may be regarded as a linking number
difference, generalized to the case of open curves. To evaluate it, we need a generalization of
the Fuller--White--Calugareanu relation 
$\Delta\Lk = \Delta\Tw + \Delta\Wr$ for open curves. We start with the 
$\star$-loop, with its untwisted frame. Next 
we construct a 
plane, untwisted, circular arc $C$, 
attached to the two ends of
 the $\star$-loop to form a  closed, smooth,
framed curve. Completing the $\star$-loop in
this way gives a closed loop with $\Tw=\Wr=0$. Also, the tangents at the ends of 
$C$ match the tangents at the ends of any of the family of loops we 
are studying. 

Completing any other loop in our family with the same $C$ gives a discontinuity in the axial
orientation at one  of the attachment points. Hence the formula
for linking number will not give an integer; instead, $2\pi\Lk$ is the 
desired formula for $\ppsi$. Setting it equal to one of the desired
values (\eref{e:psic}) gives our fourth boundary condition.

In summary, our fourth boundary 
condition is $\ppsi =(\pspacing-\pspacing_0)\pono+2\pi \ploopn$ for an integer $\ploopn $,
where \cite{kami02a}
\begin{eqnarray}\pnlabel{eq:bc10}
 \ppsi = 2\pi \Lk & =&2\pi(\Tw + \Wr) \nonumber \\ 
 & =& \int_{-\frac{L}{2}}^{\frac{L}{2}}
\kappa_{3}\dd s 
      + \frac{1}{2}\oint\oint \hat{\mathbf{t}}(s)\cdot
      \hat{\mathbf{t}}(s')\times
      \frac{\mathbf{P}(s)-\mathbf{P}(s')}{\|\mathbf{P}(
s)-\mathbf{P}(s')\|^{3}}
      \,\dd s\, \dd s'. 
\end{eqnarray}
Here $\oint \ldots\dd s$ refers to a line integral over the closed loop
consisting of the arc $C$ plus the open loop under study.

\subsubsection{Solution strategy\pnlabel{s:ss}}
There are four parameters in the above equations: $\pz_{1}$, $\pz_{2}$, $\pz_{
3}$
and $\lambda$. As in \sref{s:plane}, we must find values for these 
parameters
by imposing the boundary conditions. These four parameters can be determined by enforcing the boundary conditions
(\erefs{eq:bc1}, \ref{eq:plane1}, \ref{eq:bc7}, and \ref{eq:bc10}). 
This leads to a set of equations that must be solved numerically, using
Newton's method. Our initial 
guess for $\pz_{1}$, $\pz_{2}$, $\pz_{3}$ and $\lambda$ for given boundary data 
corresponds to the values of these parameters for a planar loop. 
For example, we know how to solve for $k$ and $\lambda$ for a 
planar loop (for which $\ppsi = 0$)  of length $L$, end-separation $a$ and 
end-angle $\theta_{a}$. For a three dimensional loop with similar data (but
$\ppsi \neq 0$) we initialize Newton's method with the guess
$\pz_{1} =-1$, $\pz_{2} = 2k^{2}-1$, $\pz_{
3} =1$. We then make a small increment
in $\ppsi$ and solve a system of four algebraic equations to obtain
the nearby values of $\pz_{1}$, $\pz_{2}$, $\pz_{3}$ 
and $\lambda$ 
that give this value
of $\ppsi$. We continue this incremental process until we have achieved
one of the 
values of $\ppsi_{\ploopn}$ dictated by the contour length $L$ between the repressor 
binding sites (\eref{e:psic}). This numerical procedure corresponds physically to holding a rod 
into a planar loop and
subsequently rotating one end, continuously changing the shape of the loop. 
Once we have computed this solution, the elastic energy stored in the DNA is 
evaluated
 using the following expression (see \cite{niz99a}):
\begin{equation}
 \pE_{\ploopn}
  = \frac{FL}{2}\Big[\pz_{1}+\pz_{2}+\pz_{3}-\pz_{1}\pz_{2}\pz_{3} 
 - \sqrt{(1-
\pz_{1}^{2})
 (1-\pz_{2}^{2})(\pz_{3}^{2}-1)}\Big] + \frac{{M_3}^{2}L}{2\pktw }.
\end{equation} 
Then we continue the incremental search both forward and backward in 
$\ppsi$ looking for other topoisomers.

\section{Results\pnlabel{s:r}}
Our goal is to capture certain broad features of the looping free energy
change as obtained from experiments in the analysis of \rrefs{saiz05a,garc?a} 
(\frefn{ftwo}b). Beyond the gross
structure,  there is a shallow minimum in
the looping free energy change, in the 70--80bp range. Keeping in mind 
that the horizontal axis of the graph differs from our $\pspacing$ by 
21$\,$bp, this minimum corresponds to $\pspacing\approx50$--60$\,$bp. 
In contrast, for loops formed in
cyclization reactions \cite{clou05b} the minimum in free energy change occurs
at about 500bp \cite{shim84a}. We will see that our elastic rod model, incorporating
supercoiling effects and a highly simplified form of the geometry of
the repressor--DNA complex, does reproduce some qualitative features in
the length dependence of the free energy change. To do this, we
now apply the methods of analysis outlined in the previous sections.

\sref{s:pf} described the idealizations we have made to keep the role
of external torsional stress as clear as possible. These idealizations 
limit our ability to make quantitative predictions for specific
systems, but nevertheless we will apply our method using
parameters partially inspired by the structure of the {\it lac}
repressor complex. Thus, we
estimate the distance between the two ends of the DNA to be $a \approx 
\pavalue\, $nm. 
We estimate  $\ptorq  = -1.0\,\kbt$ (see \sref{s:es}), and take effective values for the
elastic constants from experiments on the cyclization of short 
DNA\pfoots\footnote{The twist stiffness given here is  
smaller than the value found in single-molecule studies. Widom and 
Cloutier found that this small value was needed to fit their data on 
cyclization of short DNA, and proposed that it was an effective value 
reflecting a nonlinear breakdown of elasticity under high strain 
\cite{clou05b}. Previous authors also found that a significant, though
less dramatic, reduction of the value of twist stiffess was needed to fit 
cyclization of longer DNA \cite{shim84a}. In the present work we found 
that a reduced effective twist stiffness was needed to get the 
required magnitude of the peak-to-valley energy change in 
\frefn{ftwo}b; however, Saiz and Vilar have argued that the existence 
of multiple looping geometries can also reduce this modulation 
\cite{saiz?a}.}: 
$\pxib  = \xibval\, $nm, $\pxitw  = \xitval\, $nm
\cite{clou05b}. Finally, we make initial guesses $\pthetaa=\pthetaaval^{\circ}$ and
$\pspacing_0=0$ for the unknown parameters.

\ifigure{fig:twisttheory}{{\it a.} The elastic energy of DNA loops as calculated from an elastic rod
  model of DNA. We have assumed 
 $\theta_{a} = 71^{\circ}$ and $a = \pavalue $nm with $\xi_{p} = 50$nm, $\xi_{t} = 18$nm 
 and $\ptorq  =-1.0\kbt  $. The graph shows the quantity defined in 
 \eref{e:DG}, plus an arbitrary linear function of 
 $\pspacing$, because such a function was dropped in our derivation of \eref{e:DG}.
 The exchange of stability between various topoisomers at the peaks of the 
 modulations (40,52,63,74bp) has been emphasized by plotting the
 energy of the two competing topoisomers with different symbols (circles 
 and stars).
 The continuous line
 connects the lowest energy topoisomers at each value of the length $L$ of the loop. 
{\it b.} The black curve plots the elastic energy of  
 a planar loop without any twist, as calculated using \eref{eq:planenerg}.
 This curve would be appropriate for looping with nicked 
 DNA.}{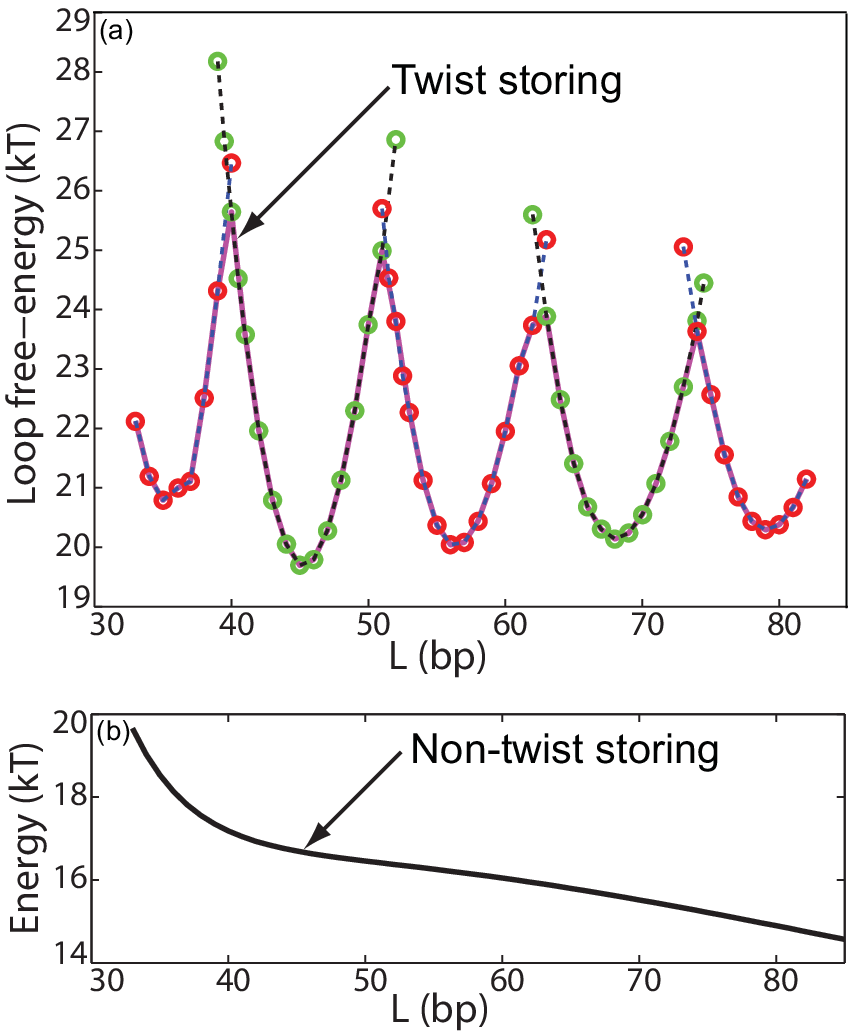}{3.9}

\ifigure{fig:predictions}{Predicting the effect of changing the sign and magnitude of $\ptorq $.
Again we took $\theta_{a} = 71^{\circ}$ and $a = \pavalue $nm with $\xi_{b} = 50$nm, 
 $\xi_{tw} = 18$nm and plotted the energy for three values of $\ptorq 
 $. All three panels have the same, arbitrary, linear function of $L$ 
 added to the looping free energy change. The sign of the assumed torque $\ptorq $ in panel (c) is the opposite of that in 
 panel (a). 
The minima and maxima in 
 panel (c) are shifted by about 4$\,$bp as compared to those in panel 
 (a).}{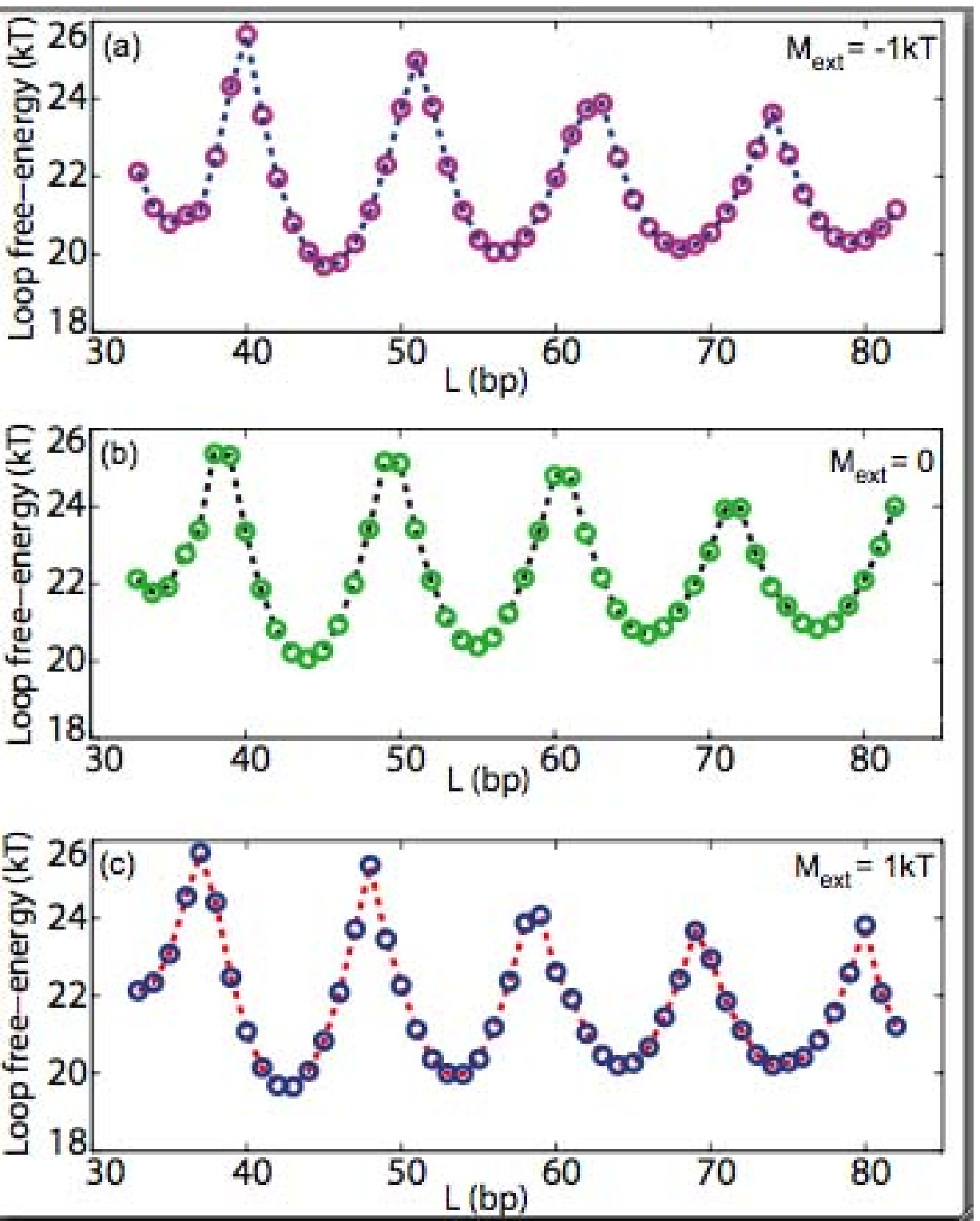}{3.9}

\fref{fig:twisttheory} shows the free energy of loops with 
$\ptorq  =-1.0\kbt  $. This result shows that a simple elastic rod model of the 
DNA is able to capture the general trend in the modulations of the free 
energy. The amplitude of the modulations (about $4 \kbt $) is correctly
reproduced and the maxima of the modulations are sharp, as found from 
experimental data by Saiz {\it et al.} \cite{saiz05a}. The curve also 
shows a dip in free energy at about 
45$\,$bp, fairly close to the dip in the experimental data. No such 
dip is seen in the free energy of looping for nicked (non-twist 
storing) DNA, so its appearance is influenced by  the external torque due to 
supercoiling. 

\fref{fig:predictions} shows the looping free energy change
as a function of length for the three values 
$\ptorq  =-1, 0,$ and $+1\kbt $, illustrating our point that the exchanges of
stability that determine the dominant topoisomer at a given value of $L$ depend
on the magnitude and sign of the external torsional stress.
This is easily seen by comparing panels (a) and (c) of 
\fref{fig:predictions} which show that  
changing the sign of $\ptorq $ shifts the maxima of the modulations
by about 4$\,$bp. The sign of $\ptorq $ can in principle be controlled in an 
{\it in vitro} experiment, such as a magnetic bead assay (employed in Lia
{\it et al.} \cite{lia03a}). 
As a result the preference for a particular topoisomer of a loop will change 
and this will result in an altered dependence of the looping free energy on 
the contour length. \fref{fig:predictions} summarizes how this dependence 
will be altered for zero torque and a positive torque. \fref{fig:predictions} 
has been constructed for the geometry of the {\it lac} repressor but the 
conclusion that the magnitude and sign of the external torque $\ptorq $ 
controls the locations of the minima and maxima of the modulations in the 
free energy of loop formation remains valid for any other DNA looping protein
as well. 
   
\section{Discussion}
We have shown in this paper that an elastic rod model of DNA can explain 
certain the features  observed in the length 
dependence of {\it in vivo} DNA looping free energy, if we account for the size 
and shape of the repressor--DNA complex, and for the external torsional stress
from supercoiling in the bacterial chromosome. These features have not been 
adequately examined in the theoretical literature, although they are critical
in determining the function of the repressor--DNA complex. 
We have also obtained predictions that could elucidate
the mechanics of protein--DNA interactions, and we hope that they will inspire
future experiments. One can adjust $\ptorq$ {in vivo} by 
studying bacteria with varying levels of supercoiling density, for 
example by disabling the topoisomerases that normally maintain the genome
under torsional stress
(\cite{volobook}). Or the present theory can be generalized to 
incorporate the effects of stretching force; then 
a magnetic bead assay could be used to test
our prediction that the locations of the minima and maxima in the modulations
of the free energy of loop formation can be 
controlled by the applied external torque. 
An important limitation of our theory, as presently stated, is that it does 
not address entropic effects and therefore is applicable only for short 
contour lengths of DNA.However, the efficient numerical approach developed 
in this paper remains applicable for longer contour lengths as well and 
could be used in conjunction with Monte Carlo methods or Molecular Dynamics 
to explore the effects of entropy on the mechanics of protein DNA interactions.

\begin{acknowledgments}
We thank John Beausang, Laura Finzi, Hernan Garcia, Jeff Gelles, 
Randall Kamien, Igor
Kulic, Wilma Olson, Rob Phillips, Leonor Saiz, Robert Schleif, Andrew Spakowitz, David Swigon, and Paul Wiggins for helpful 
discussions. PN and PP acknowledge the Human Frontier Science 
Program Organization for partial support, and
PN acknowledges NSF Grant DMR-0404674 and the Nano/Bio Interface
Center, National Science Foundation DMR04-25780. PN acknowledges
the hospitality of the Kavli Institute for 
Theoretical Physics, supported
in part by the National Science Foundation under 
Grant PHY99-07949,
where some of this work was done.

\end{acknowledgments}

\newpage\appendix
\section{Summary of notation\pnlabel{s:sn}}
\pitemize
\item
$\pkb$, $\pktw$ are the bend and torsional elastic constants of DNA; the 
corresponding persistence lengths are $\pxib$ and $\pxitw$.\\
$\pono$ is the natural twist rate of DNA, about $2\pi$ radians per 
11bp. $\phelix$ is the helix pitch of DNA, about 11bp.

\item
$s\pini$, $s\pfin$ are the arclength positions at which the DNA exits its
binding sites and enters the loop interior; $\tilde s\pini$, $\tilde s\pfin$ are the 
corresponding positions where the DNA enters the exterior region.
\\
$\pspacing=s\pini-s\pfin$ is the spacing between exit points; $\pspacing_0$ is a special
value of this spacing for which the coupler admits a planar, untwisted 
loop.
\item
$\{\ehat_i(s), i=1,2,3\}$ denote the physical orthonormal frame attached to 
the DNA at arclength position
$s$; $\{\dhat_i(s)\}$ is the corresponding untwisted frame.\\
$\kappa_{1,2,3}(s)$ denote components of the curvature vector at location 
$s$, measured in the untwisted frame.
\\
$\pthat=\ehat_3=\dhat_3$ is the tangent vector to the DNA centerline, as a function of
arclength position along the DNA.

\item 
The 
``coupler'' refers to a geometrical representation of a regulatory 
protein complex, imposing a fixed relation between the spatial
locations and physical orientations of two points on the DNA. It is
independent of the spacing $\pspacing$. A ``physical loop configuration'' is
one obeying the boundary conditions imposed by the coupler.
\\
The ``\starc'' is a fictitious, modified coupler differing from the
physical one by an $\pspacing$-dependent axial rotation of one of the
DNA strands relative to the other. Quantities associated to it are
denoted by a subscript $\star$. The ``$\star$-loop configuration'' is
the loop of minimal elastic energy obeying the boundary conditions imposed by the 
\starc.
\\
$\pavec$ is the spatial separation between DNA detachment points;
$\pasca$ denotes its length.
\\
$\pthetaa$ is the exit angle characterizing the regulatory
protein complex.
\\
$\beta\qquad$ twist angle of the DNA--protein complex, set equal to
zero in this paper
\item 
$\ptorq\qquad$ torsional stress outside the looping region, same units 
as energy; $M_{1,2,3}(s)$ are the components of the moment (torque) vector 
inside the loop at
arclength position $s$, expressed in the untwisted frame. Note that 
$\ptorq\not=M_3$
  in general.
\item 
$\pE\st$ denotes the elastic energy of the lowest-energy $\star$-loop
configuration; $\pE$ denotes the elastic energy of a physical looped state.
\\
$\ploopn$ indexes which of the possible physical loop states is under 
discussion; $\pulst$ is the unlooped state.
\\
$\ppsi_\ploopn$ is the axial rotation angle by which physical loop 
$\ploopn$ differs  from the $\star$-loop.
\item 
$G\puns(\pLtot,\sigma)\qquad$ denotes the free energy of an unlooped
circular DNA of length $\pLtot$ with fractional excess linking number
$\sigma$; $\mu(\sigma)$ is the corresponding free energy per unit
length.
\item 
$\am,\sn,\cn$ are the usual elliptic functions. $k$ is the
modulus of an elliptic function.
$E(y|k)$ is the incomplete elliptic integral of the second kind;
$\Pi(y|n,k)$ is the elliptic function of the third 
kind.

\item
$\pz_{1,2,3}$ are parameters entering the general elastic 
equilibrium solution in 3D.
\\
$R(s), \Phi(s), Z(s)$ are cylindrical coordinates for the position 
of the rod at arclength position $s$.
\\
$\theta(s),\phi(s)$ are spherical polar coordinates for the unit 
tangent vector to the rod at position $s$.
\xpitemize

\bibliographystyle{prsty}
\bibliography{purohit}
\end{document}
\newpage
FIGURE LABELS

$\tilde s\pini$

$\tilde s\pfin$

$ s\pini$

$s\pfin$

$\pthat\pini$

$\pthat\pfin$

$\pthetaa$

$\pavec$

$\plst$

$\plst'$

$\pulst$

$s$

$y$

$z$

$\pnvec F$

$\pavec$

\end{document}